\documentclass{article}
\usepackage{graphicx} % Required for inserting images
\usepackage{amsmath}
\usepackage{caption}
\usepackage{subcaption}
\usepackage{enumitem}
\usepackage{url} 
\usepackage{hyperref} 
\title{A3C3: AI Algorithm and Accelerator Co-design, Co-search, and Co-generation}
\author{
Selin Yildirim$^{1}$ \and
Yingbing Huang$^{2}$ \and
Deming Chen$^{2}$ \\[0.75em]
$^{1}$Siebel School of Computing and Data Science \\
$^{2}$Department of Electrical and Computer Engineering \\
University of Illinois Urbana-Champaign \\[0.75em]
\texttt{\{seliny2,\;yh21,\;dchen\}@illinois.edu}
}

\date{February 2026}

\begin{document}

\maketitle

\section{Abstract}

We present a holistic methodology for artificial intelligence algorithm and accelerator co-design, co-search, and co-generation (A3C3), which jointly optimizes neural network architectures and their hardware implementations to address the inefficiencies of traditional top-down AI system design flows. Conventional AI deployment often treats model design and hardware mapping as separate stages: an algorithm is first developed for accuracy, and only afterward adapted to meet latency, throughput, energy, or resource constraints. This separation can lead to suboptimal systems, particularly as modern AI workloads become increasingly heterogeneous, memory-intensive, and platform-dependent. A3C3 instead parameterizes both algorithmic and accelerator design spaces and searches them jointly, enabling the automatic generation of model–accelerator pairs that better balance accuracy, latency, throughput, energy efficiency, and hardware utilization.

A key component of this methodology is differentiable architecture and implementation co-search, which formulates design decisions as continuous, optimizable variables rather than relying solely on discrete manual exploration. In this setting, neural architecture choices, hardware mapping strategies, and implementation parameters can be optimized using gradient-based methods under hardware-aware objectives. For example, the efficient differentiable deep neural network architecture and implementation co-search framework \cite{li2020edd} demonstrates how model structures and deployment configurations can be jointly optimized across graphics processing units (GPUs) and field-programmable gate arrays (FPGAs). By incorporating hardware cost models directly into the search process, differentiable co-search enables more systematic exploration of large design spaces and produces implementations that are not only accurate but also efficient on the target platform.

Beyond conventional neural network acceleration, this chapter also examines how A3C3 principles extend to large language models (LLMs), where inference efficiency is increasingly limited by autoregressive decoding and memory movement. Speculative decoding~\cite{leviathan2023fast, chen2023accelerating} accelerates LLM generation by using a lightweight and auxiliary prediction mechanism (i.e, draft model) to propose multiple future tokens, which are then verified by the larger target model in parallel. This approach increases decoding speed while preserving the output distribution of the target model via rejection sampling, where draft tokens are accepted only when they are statistically consistent with the target model and otherwise resampled from the target distribution. Medusa~\cite{cai2024medusa} further advances this direction by attaching multiple decoding heads to the target model, allowing it to generate and verify candidate sequences more efficiently without requiring a separate draft model. These speculative decoding architectures illustrate how algorithmic restructuring and hardware-conscious execution can jointly reduce latency and improve throughput for large-scale generative AI inference.

The chapter demonstrates A3C3 through multiple representative systems, including SkyNet~\cite{zhang2020skynet} for real-time edge vision, differentiable architecture and implementation co-search~\cite{li2020edd} for GPU and FPGA deployment, speculative decoding methods for LLM acceleration~\cite{leviathan2023fast, chen2023accelerating,cai2024medusa}, and dynamic Key-Value (KV) cache compression through SnapKV~\cite{li2024snapkv}. Across diverse platforms and application domains, these systems show substantial gains in performance, efficiency, and scalability. Collectively, this body of work establishes A3C3 as an effective and extensible foundation for future adaptive, heterogeneous, and large-scale AI systems, positioning co-design as a central paradigm for next-generation AI and hardware innovation.

\section{Introduction}
The success of deep learning over the past decade has fundamentally reshaped modern computing systems. Convolutional neural networks (CNNs)~\cite{o2015introduction}, recurrent neural networks (RNNs)~\cite{schmidt2019recurrent} and Transformer~\cite{vaswani2017attention} models have achieved unprecedented accuracy across vision, speech, natural language processing and decision-making tasks. However, these algorithmic advances have been accompanied by dramatic increases in computational complexity, memory footprint, and energy consumption. As a result, deploying state-of-the-art models on real hardware, particularly under tight latency, power, and cost constraints, has become increasingly challenging. 

Traditional AI system development largely follows a top-down design paradigm. In this workflow, neural network architectures are first designed and trained with minimal consideration of hardware constraints, often prioritizing accuracy on standardized datasets. Only after the model has been finalized is it mapped onto a target hardware platform, such as a GPU, FPGA, or Application-Specific Integrated Circuit (ASIC) accelerator. While this separation of concerns simplifies early-stage algorithm development, it frequently leads to sub-optimal system-level outcomes, especially in resource-constrained or real-time settings. 

Several fundamental limitations characterize this top-down approach. First, software-level metrics (accuracy, robustness, generalization) and hardware-level metrics (latency, throughput, energy, silicon area) are optimized independently, making it difficult to reason about trade-offs across domains. Second, the choice of a reference neural architecture at the outset is often guided by empirical experience or benchmark popularity rather than by hardware efficiency. Third, deployment typically requires long and iterative cycles of manual tuning, involving quantization \cite{jacob2018quantization}, pruning, tiling, and scheduling, which are both time-consuming and error-prone. These issues are particularly pronounced for embedded and edge platforms, where performance margins are narrow and design errors are costly. 

\begin{figure}[!hbtp]
    \centering    \includegraphics[width=\linewidth]{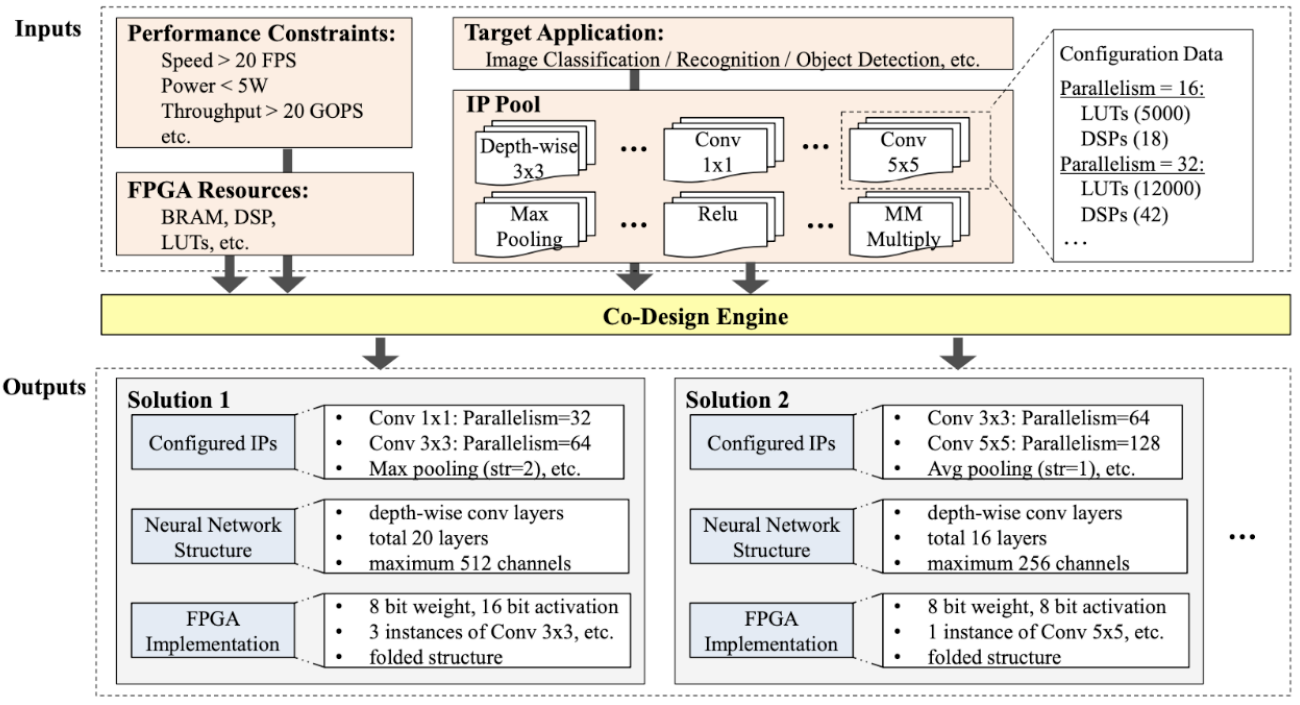}
    \caption{Proposed A3C3 Method in ICSICT 2018~\cite{hao2018deep}}
    \label{fig:icsict}
\end{figure}

To address these challenges, AI Algorithm and Accelerator Co-design, Co-search, and Co-generation is proposed as a fundamentally different methodology. Instead of treating neural architectures and hardware implementations as separate artifacts, A3C3 views them as a coupled design problem. As shown in Figure \ref{fig:icsict}, the central idea is to jointly parameterize, explore, and optimize both the algorithmic and hardware design spaces, and to automatically generate matched model–accelerator pairs that meet application-specific requirements. 

This chapter consolidates a series of research efforts that progressively develop, refine, and extend the A3C3 methodology. Beginning with early co-design concepts for FPGA-based CNN accelerators~\cite{hao2019fpga}, the work evolves into scalable frameworks such as SkyNet \cite{zhang2020skynet} and EDD \cite{li2020edd}, and ultimately expand to address the emerging challenges of large language model (LLM) inference. Together, these systems illustrate how co-design can serve as a unifying principle across diverse AI workloads and hardware platforms. 

\section{The A3C3 Methodology}
A central challenge in deploying neural networks on resource-constrained hardware is the sheer complexity of the hardware design process itself. Tools such as high-level synthesis (HLS) \cite{chen2005xpilot, papakonstantinou2009fcuda, cong2022fpga, ye2022scalehls} compilers have attempted to bridge this gap by automatically translating algorithmic descriptions into hardware implementations. For instance, the hierarchical dataflow compiler for high-level synthesis (HIDA) \cite{ye2024hida} demonstrates that even the relatively narrow problem of dataflow optimization on FPGAs harbors a combinatorially explosive, inter-task-coupled design space, one where decisions around loop tiling, spatial parallelization factors, on-chip buffer partitioning, and streaming channel sizing are tightly co-dependent. HIDA addresses this through a two-tier intermediate representation operating at both functional and structural abstraction layers, combined with a pattern-driven task fusion and intensity-aware parallelization optimizer, achieving up to 8.54x higher throughput over prior HLS tools while reducing design time from hundreds of engineering hours to under 10 minutes. Yet HIDA, like all HLS-centric approaches, optimizes hardware implementation in isolation. It accepts a fixed neural architecture as input and asks only: how do we best map this to silicon? This framing, while powerful, leaves a deeper question unaddressed: what if the architecture itself is wrong for the hardware? This is precisely the gap that the A3C3 methodology is designed to close. A3C3 is built on the premise that optimal AI systems cannot be obtained by optimizing algorithms or hardware in isolation.

\subsection{Design Philosophy and Scope} 
\begin{figure}[htbp]
    \centering
    \includegraphics[width=0.6\linewidth]{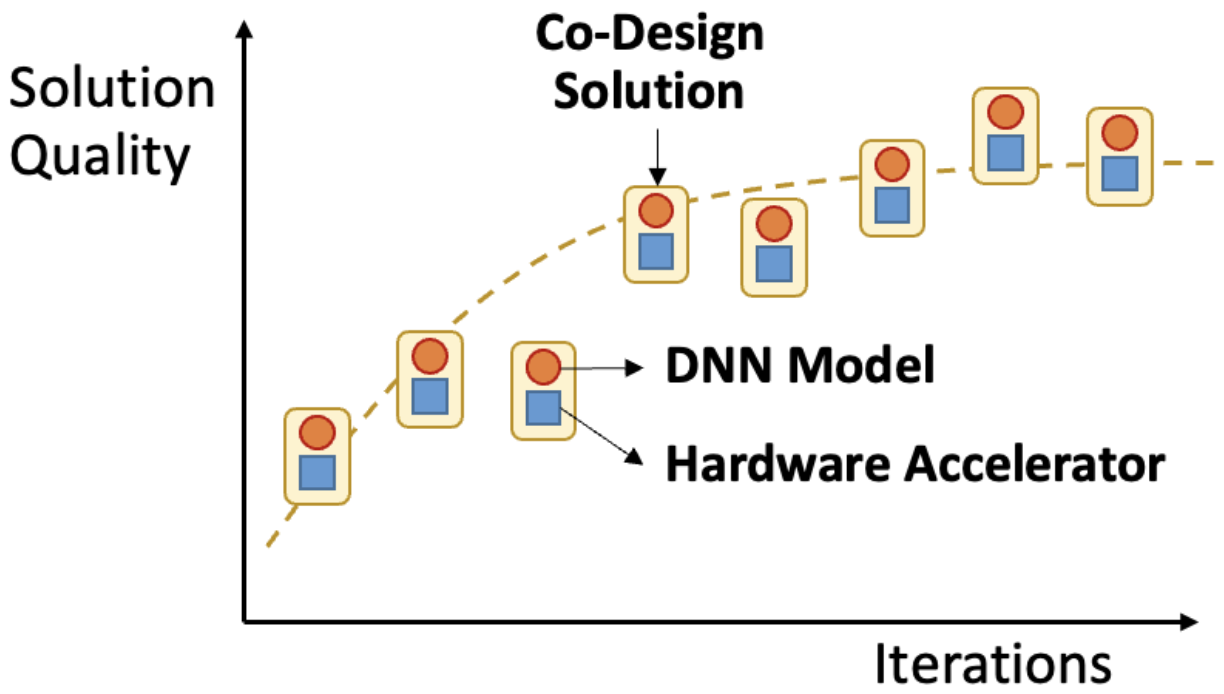}
    \caption{Iterative co-search of DNN models and hardware accelerators in A3C3 design. Each iteration evaluates a candidate model-accelerator pair, progressively improving solution quality under QoS and QoR constraints.}
    \label{fig:iter}
\end{figure}

A3C3 is built on the premise that optimal AI systems cannot be obtained by optimizing algorithms or hardware in isolation. Instead, system-level optimality emerges only when the interactions between neural architectures, dataflows, memory hierarchies, and compute resources are explicitly modeled and optimized together. Accordingly, A3C3 defines a joint design space that spans: 
\begin{itemize}
    \item \textbf{Algorithmic parameters:} This includes layer types (e.g., $3\times3$ convolutions, $1\times1$ pointwise convolutions, depth-wise convolutions), kernel sizes, channel widths, network depth, and quantization levels (e.g., 8-bit weight/16-bit activation or 8-bit/8-bit configurations).
    \item \textbf{Hardware parameters:} This involves degrees of parallelism, loop tiling factors, data reuse strategies, and memory partitioning. 
    \item \textbf{System-level objectives:} Objectives go beyond accuracy to include latency, throughput such as Giga Operations per Second (GOPS), energy consumption, and utilization of FPGA resources such as Block Random Access Memory (BRAM), Dedicated Silicon Blocks (DSB), Look-Up Tables (LUT). Accordingly, optimizations are guided by hard thresholds such as speed ($> 20$ frames per second), power consumption ($<5$ Watt), and throughput ($> 20$ GOPS).
\end{itemize}
The goal of A3C3 is to avoid the sub-optimality of traditional top-down AI system design, where a model is first developed for algorithmic accuracy and only later forced onto a hardware platform through compression, pruning, quantization, or manual mapping. Such a sequential flow often produces hardware-unfriendly models whose computation patterns, memory access behavior, and resource requirements are poorly matched to the target accelerator. A3C3 instead treats the deep neural network (DNN) model and hardware accelerator as a coupled design pair. As shown in Figure~\ref{fig:iter}, the model architecture and accelerator implementation are iteratively searched together, with each design iteration improving the overall solution quality. The dashed curve illustrates the progressive improvement obtained through co-search, while each icon represents a candidate model-accelerator pair evaluated under application-level quality-of-service (QoS) and quality-of-result (QoR) constraints. Rather than optimizing the model or hardware in isolation, A3C3 seeks a co-design solution in which the algorithm is hardware-friendly by construction and the accelerator is specialized to the model’s computational structure.
\subsection{Bundles as Unified Building Blocks} The foundational abstraction within the A3C3 methodology is the \textit{bundle}. Unlike traditional design paradigms that treat neural network layers and hardware kernels as disparate entities, a bundle serves as a unified building block that encapsulates both a sequence of mathematical operations and their physical hardware implementation. 
\begin{figure}[htbp]
    \centering
    \begin{subfigure}[b]{0.6\linewidth}
        \centering
        \includegraphics[width=\linewidth]{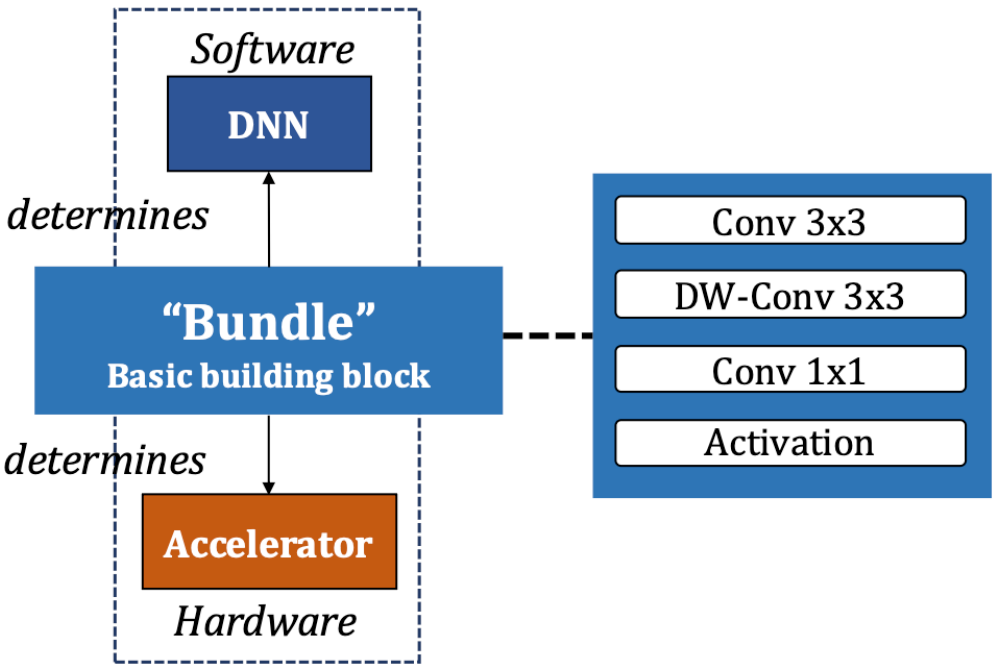}
        \caption{Role of a Bundle in A3C3 design}
        \label{fig:bundle}
    \end{subfigure}
    \hfill
    \begin{subfigure}[b]{0.35\linewidth}
        \centering
        \includegraphics[width=\linewidth]{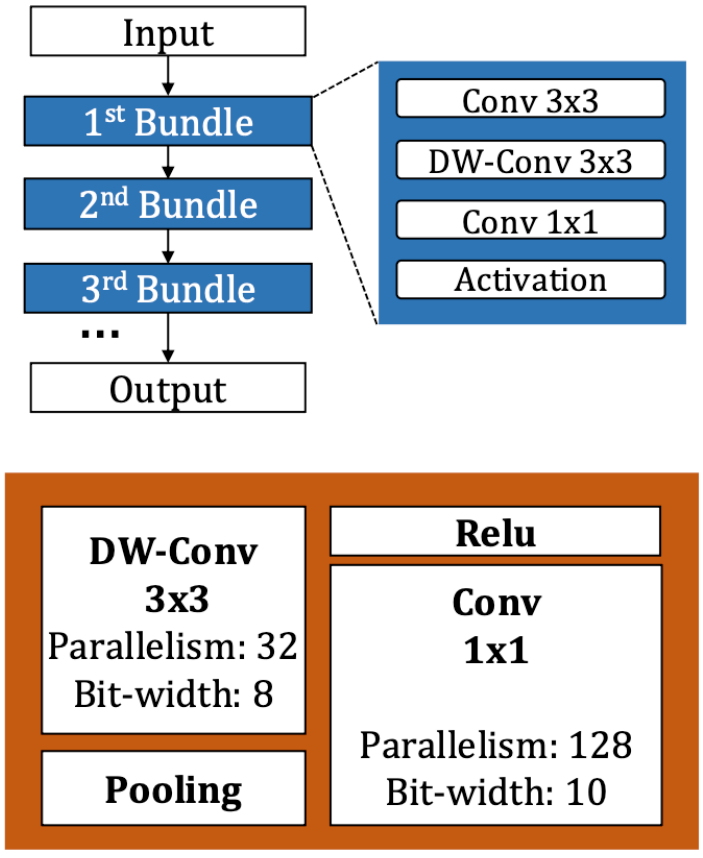}
        \caption{Stackable bundles in A3C3 design}
        \label{fig:bundle-stack}
    \end{subfigure}
    \caption{Illustration of bundle usage in A3C3 design.}
    \label{fig:bundles}
\end{figure}
Structurally, a bundle defines a specific computational subgraph, as shown in Figure \ref{fig:bundle}. For instance, a $3\times3$ convolution followed by a $1\times1$ pointwise convolution and a depthwise activation, coupled with a set of deterministic hardware parameters; such as loop tiling factors, parallelism degrees, and bit-widths. This dual-representation ensures that every algorithmic choice is grounded in physical reality. The utility of the bundle abstraction is three-fold:
\begin{enumerate}
\item \textbf{Joint Space Dimensionality Reduction:} By grouping frequently co-occurring operations (e.g., Conv $3\times3 \rightarrow$ Depth-Wise Conv $3\times3 \rightarrow$ Conv $1\times1 \rightarrow$ Activation) into a single manageable unit, bundles prune the search space. This prevents the combinatorial explosion that typically occurs when exploring neural architectures and hardware configurations independently.
\item \textbf{Hardware-Software Co-binding:} Bundles enforce a strict mapping between software layers and hardware intellectual property (IP) instances. This ensures that architectural decisions, such as a change in quantization level, are immediately and accurately reflected in hardware metrics like logic utilization and memory bandwidth. This tight coupling eliminates the risk of generating high-accuracy models that are functionally undeployable due to hardware bottlenecks.
\item \textbf{Modular Scalability:} Bundles enable a "lego-like" composition of AI systems, as demonstrated in Figure \ref{fig:bundle-stack}. Complex, high-performance networks such as SkyNet \cite{zhang2020skynet} can be rapidly constructed by stacking and configuring these pre-evaluated, hardware-verified components. This modularity facilitates the reuse of optimized Intellectual Property (IP) cores across different network topologies, significantly accelerating the design cycle from concept to deployment.
\end{enumerate}

\subsection{Co-search and Co-generation Flow}
% \begin{enumerate}
%     \item \textbf{Bundle Selection:} A library of candidate bundles is defined, capturing a range of algorithmic and hardware trade-offs.
%     \item \textbf{Joint Architecture Exploration:} Neural networks and accelerators are constructed by composing bundles according to predefined templates, while exploring variations in structure and implementation. 
%     \item \textbf{Refinement and Deployment:} The selected model–accelerator pair is fine-tuned, quantized, and deployed on the target platform. 
% \end{enumerate} 

The A3C3 workflow is designed to overcome the limitations of sequential design by treating the neural architecture and the hardware accelerator as a single, coupled entity. This process moves beyond simple heuristic tuning, employing a structured three-stage pipeline that automates the transition from abstract algorithmic requirements to physical hardware implementation. The A3C3 workflow proceeds in the following three stages.
\subsubsection{Stage 1: Multi-Objective Bundle Evaluation} The flow begins with the population of a bundle library. Unlike traditional libraries that only list software kernels, A3C3 bundles are pre-characterized for specific target hardware (e.g., FPGAs, GPUs, or ASICs). During this stage, each bundle is evaluated across multiple dimensions:
\begin{itemize}
\item \textbf{Algorithmic Efficacy:} Measured via accuracy (e.g, Intersection over Union (IoU)) on a validation subset for image object detection.
\item \textbf{Hardware Efficiency:} Measured via cycles per iteration, GOPS, power consumption etc.
\item \textbf{Resource Footprint:} Quantifying the utilization of physical units such as DSPs, BRAMs, and LUTs.
\end{itemize}
By plotting these metrics, the system generates a Pareto-optimal frontier, allowing the co-search engine to discard sub-optimal "hardware-unfriendly" blocks before the global search begins.

\subsubsection{Stage 2: Joint Search Space Exploration}

Once the library is established, the methodology enters the joint exploration phase. This stage utilizes a search engine, often based on evolutionary algorithms, reinforcement learning, or gradient-based methods, to navigate a high-dimensional design space. The search is governed by a joint objective function, which mathematically represents the trade-offs between competing requirements. Specifically, it navigates a high-dimensional design space where the algorithm ($A$) and the implementation ($I$) are interdependent variables and are explored simultaneously. For example, the search might decide to increase the depth of the network on the algorithm side while simultaneously increasing the loop-unrolling factor, parallelism level, or tiling configuration on the accelerator side to ensure that the latency constraint is still satisfied.

A representative example of this process is the efficient differentiable DNN architecture and implementation co-search framework (EDD)~\cite{li2020edd}, which formulates the joint search problem over a fused design space $\{A,I\}$. In EDD, the architecture space includes candidate neural operations within each DNN block, while the implementation space includes hardware-dependent choices such as quantization precision and platform-specific mapping parameters. Rather than selecting an architecture first and optimizing its hardware implementation afterward, EDD relaxes both types of discrete decisions into differentiable variables and optimizes them jointly with a hardware-aware loss. The objective accounts for task accuracy, performance metrics such as latency, throughput, energy, or model size, and resource constraints imposed by the target platform. This illustrates the central idea of Stage~2: the search engine does not merely find the most accurate model or the fastest accelerator in isolation, but instead identifies a feasible algorithm-implementation pair whose combined behavior satisfies the target QoS and QoR requirements.
\subsubsection{Stage 3: Automated Co-generation and Synthesis}
The final stage of the flow is the automated translation of the optimal design pair into deployable artifacts. This involves a bifurcated generation process:
\begin{enumerate}
\item \textbf{Software Generation:} The flow generates the final model topology and trained weights. It often includes hardware-aware refinement steps, such as fine-grained quantization or channel shuffling and expansion, to maximize the efficiency of the underlying hardware dataflow.
\item \textbf{Hardware Synthesis:} The flow generates synthesizable hardware description code (e.g., Verilog or High-Level-Synthesis-based C++). This includes the configuration of memory hierarchies, communication, and processing elements (PEs) specifically tailored to the data reuse patterns of the selected neural architecture.
\end{enumerate}
By automating this co-generation, A3C3 ensures that the final AI system is not just a software model running on generic hardware, but a bespoke, co-designed pair where the hardware is specialized for the algorithm, and the algorithm is tuned for the hardware’s physical constraints. Thus, the output of A3C3 process is a fully specified AI system which includes both the trained model and the hardware configuration required to execute it efficiently.

The following sections instantiate the A3C3 paradigm across progressively broader AI system settings, beginning with edge vision and extending toward differentiable co-search and large-scale language-model inference. SkyNet first demonstrates how algorithm-accelerator co-design can produce compact, hardware-friendly convolutional networks for real-time UAV object detection, where small-object detection accuracy must be preserved under strict power, latency, and resource constraints. EDD then generalizes this idea by replacing coarse heuristic exploration with differentiable architecture and implementation co-search, allowing neural operations, quantization choices, and hardware mapping parameters to be optimized within a unified design space. The chapter then shifts to large language models, where the dominant bottlenecks are no longer only convolutional compute or embedded resource usage, but autoregressive decoding latency, memory bandwidth, and KV-cache growth. Medusa addresses this regime by co-designing speculative decoding structures with GPU execution characteristics, while SnapKV targets long-context inference by dynamically compressing the KV cache based on input-dependent attention behavior. Together, these case studies show that A3C3 is not tied to a single model family or accelerator type; rather, it provides a general methodology for identifying the dominant system bottleneck and jointly adapting the algorithm, memory behavior, and hardware implementation to satisfy application-level QoS and QoR requirements.
\section{SkyNet: Co-designed Edge Vision Systems}
\begin{figure}[!hbtp]
    \centering
    \includegraphics[width=\linewidth]{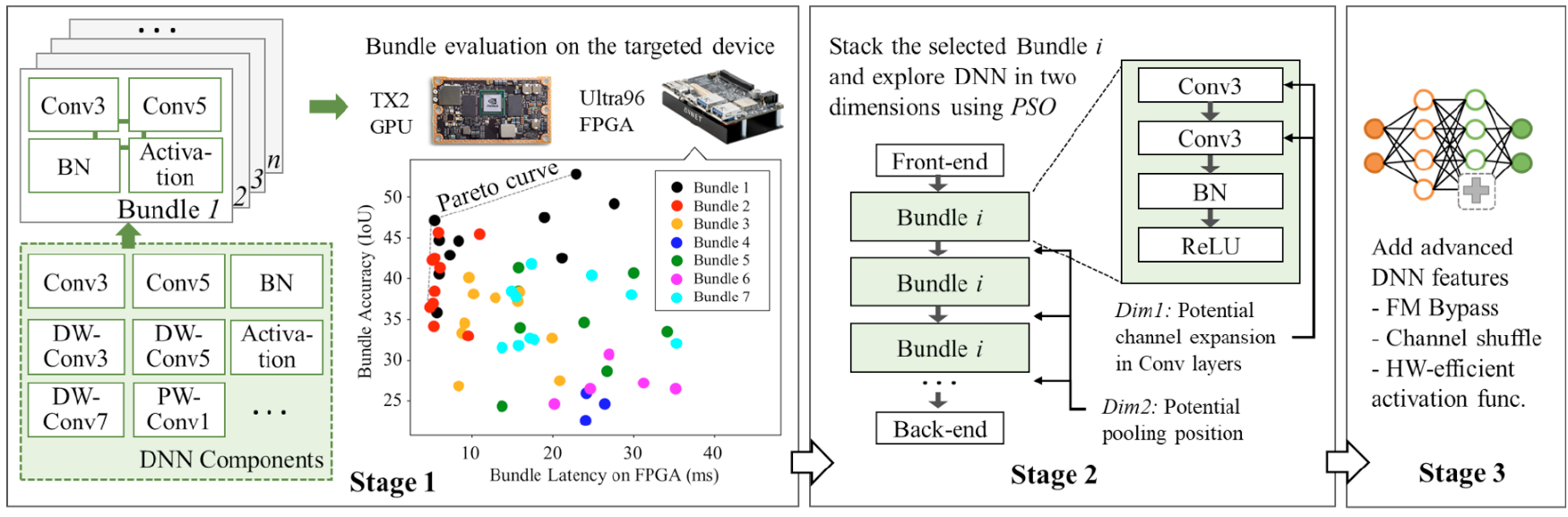}
    \caption{The A3C3 workflow proposed by SkyNet \cite{zhang2020skynet}, a hardware-efficient method for object detection and tracking on embedded systems. }
    \label{fig:a3c3}
\end{figure}
\subsection{Application Context}
SkyNet \cite{zhang2020skynet} is developed as a concrete instantiation of the A3C3 methodology for real-time object detection on unmanned aerial vehicle (UAV) platforms. This domain is characterized by stringent constraints on power, latency, and weight, often requiring high-frame-rate processing on embedded FPGAs or GPUs. The target application, motivated by a system design competition \cite{xu2019dac}, involves detecting objects across 95 categories. A significant challenge is the scale of the targets: many objects (31\% of the targets) occupy less than 1\% of the total image area, while 91\% of the targets occupy only 9\% of the area. This necessitating a network that maintains high spatial resolution without incurring the computational cost of traditional heavy backbones.
\subsection{Architecture and Accelerator Co-design}
The SkyNet neural architecture consists of 13 convolutional layers with approximately 0.4 million parameters. Rather than starting from an existing backbone like MobileNet~\cite{howard2019searching} or ResNet~\cite{he2016deep}, the network structure is generated through a multi-stage A3C3-based co-search. Figure~\ref{fig:a3c3} illustrates the three-stage SkyNet design flow, which applies A3C3 to jointly construct a compact DNN and a hardware-efficient accelerator for edge vision. In Stage~1, a library of candidate DNN building blocks, or bundles, is constructed from convolutional, depthwise convolutional, pointwise convolutional, batch-normalization, and activation components. These bundles are evaluated directly on the target devices, such as the TX2 GPU and Ultra96 FPGA, to characterize their accuracy-latency trade-offs and identify Pareto-efficient candidates. In Stage~2, the selected bundle is stacked to form the backbone network, and Particle Swarm Optimization (PSO) is used to explore the architecture in two dimensions: channel expansion in convolution layers and pooling-layer placement. This search is critical because channel expansion controls representational capacity and hardware cost, while pooling placement determines how much spatial resolution is preserved for small-object detection. In Stage~3, additional hardware-aware DNN features, including feature-map bypass, channel shuffle, and hardware-efficient activation functions, are introduced to improve information flow and reduce implementation overhead. Overall, the figure shows how SkyNet avoids choosing a model architecture independently from the deployment platform; instead, it evaluates candidate neural components on real hardware, searches the network structure under device constraints, and then adds accelerator-friendly architectural features to produce a model-hardware pair optimized for both detection quality and edge inference efficiency.
\subsubsection{Two-Dimension Joint Exploration}
The architecture search is conducted using Particle Swarm Optimization (PSO), exploring two critical dimensions of the design space:
\begin{itemize}
\item \textbf{Channel Expansion:} The search engine determines the optimal number of filters for each layer, balancing the richness of feature extraction against the memory bandwidth required for feature maps.
\item \textbf{Pooling Positions:} To preserve spatial information for small object detection~\cite{redmon2016you}, the placement of pooling layers is optimized. The resulting architecture utilizes only three pooling layers, maintaining a relatively high-resolution feature map ($40\times24$) deep into the network.
\end{itemize}
\subsubsection{Hardware-Software Features} To maximize throughput on FPGA and GPU platforms, SkyNet incorporates several specialized features:
\begin{itemize}
\item \textbf{Bundle-based Layers:} The design utilizes a combination of bundles ($3\times3$ Depthwise Convolutions followed by $1\times1$ Pointwise Convolutions) to reduce floating point operations (FLOPs) while maintaining a large receptive field.
\item \textbf{Feature Map (FM) Bypass and Channel Shuffle:} Inspired by ShuffleNet~\cite{zhang2018shufflenet}, these features allow for better information flow across channels without increasing the number of MAC (Multiply-Accumulate) operations.
\item \textbf{Re-organized Activation:} SkyNet employs hardware-efficient activation functions that avoid complex transcendental computations, facilitating faster implementation in fixed-point FPGA logic.
\end{itemize}
\subsubsection{FPGA Accelerator Implementation}On the hardware side, the SkyNet accelerator on the Ultra96 FPGA utilizes a highly optimized dataflow. Key strategies include:
\begin{itemize}
\item \textbf{Tiling and Data Reuse:} Large feature maps are partitioned into tiles to fit within local BRAM, maximizing on-chip data reuse and minimizing energy-intensive off-chip DDR accesses.
\item \textbf{Pipelined PEs:} Processing Elements (PEs) are configured with specific parallelism factors (e.g., searching for optimal factors like 16 or 32) to match the throughput of the depthwise and pointwise layers.
\item \textbf{Quantization:} The system employs an 8-bit weight and 16-bit activation (or 8-bit/8-bit) scheme, which provides a 4$\times$ reduction in memory footprint compared to single-precision floating point with negligible loss in Intersection over Union accuracy.
\end{itemize}
\subsection{Experimental Results} Evaluated on the DAC-SDC hidden test set, SkyNet achieved the highest overall score in each of the GPU and FPGA tracks. On the Ultra96 FPGA, SkyNet reached 0.716 IoU and 25.05 FPS at 7.26 Watt, improving IoU by 0.101 absolute points over the next-best 2019 FPGA entry while achieving the highest total score. On the TX2 embedded GPU, SkyNet delivered 0.731 IoU and 67.33 FPS, outperforming other GPU-track competitors in accuracy, throughput, and total score. Beyond object detection, SkyNet is evaluated on the GOT-10k object tracking benchmark by replacing ResNet-50 backbones in SiamRPN++ and SiamMask. With SkyNet, SiamRPN++ achieved nearly identical tracking quality to ResNet-50 while running 1.6x faster, and SiamMask achieved better tracking metrics with a 1.7x speedup, indicating that the co-designed backbone can generalize beyond detection to tracking workloads.
\subsection{Future Directions}
Despite its success, the original SkyNet design relies on a relatively coarse-grained co-search strategy, where bundle selection and architecture search are somewhat decoupled. This observation highlights the potential for unified and differentiable co-design, where hardware implementation parameters and neural architecture are optimized in a single gradient-based pass. This evolution from heuristic-based search (e.g., Particle Swarm Optimization) to differentiable AI model and hardware co-search forms the basis for subsequent methodologies like efficient differentiable deep neural networks (EDD). 

\section{EDD: Efficient Differentiable Architecture and Implementation Co-search}
\subsection{Motivation: Moving Beyond Greedy Search}
While initial co-design efforts like SkyNet successfully demonstrated the value of joint optimization, they relied on a coarse-grained, three-stage process involving heuristic bundle selection and Particle Swarm Optimization (PSO). Such approaches are often "greedy", exploring only a limited portion of the combinatorial design space. The EDD framework is introduced to solve this by forming a merged and holistic design space where both the neural architecture and the physical hardware implementation are formulated as differentiable variables, solvable via gradient descent.
\subsection{Merged Differentiable Design Space}
The core innovation of EDD is the unification of the Neural Architecture Search Space $\{A\}$ and the Implementation Search Space $\{I\}$. 
\begin{enumerate}
\item \textbf{The Co-design Objective Function:} In traditional neural architecture search (NAS)~\cite{zoph2016neural, cai2018proxylessnas, tan2019mnasnet, howard2019searching}, the implementation $I_0$ is fixed, and the optimizer only seeks to minimize accuracy loss:
\begin{equation}
\min: \mathcal{L} = \text{Acc}_{loss}(A) + \alpha \cdot \text{Perf}_{loss}(I_0) \quad \text{}
\end{equation}
EDD reformulates this into a joint optimization problem where implementation $I$ is also a variable:
\begin{equation}
\min_{A, I} : \mathcal{L} = \text{Acc}_{loss}(A, I) \cdot \text{Perf}_{loss}(I) + \beta \cdot C^{\text{RES}(I) - \text{RES}_{ub}} \quad \text{}
\end{equation}
Where,
    \begin{itemize}
        \item $\text{Acc}_{loss}(A, I)$: Captures accuracy degradation as a function of both the model topology ($A$) and hardware-related effects ($I$), such as quantization bit-widths,
        \item $\text{Perf}_{loss}(I)$: Models hardware performance metrics (latency or throughput),
        \item $\beta \cdot C^{\text{RES}(I) - \text{RES}_{ub}}$: A penalty term that enforces hardware resource budgets (e.g., BRAM, DSP, LUTs) by penalizing designs that exceed the device's physical upper bound $RES_{ub}$.
    \end{itemize}
\item \textbf{Differentiable Parameterization:} EDD employs a three-stage differentiable flow to navigate this space:
\begin{itemize}
    \item \textbf{Architectural Parameters ($\theta$):} Each block in the deep neural network (DNN) is represented as a weighted combination of candidate operations (e.g., $3\times3$ or $5\times5$ depthwise convolutions, $1\times1$ convolutions). The sampling parameters $\theta$ are optimized to select the best operation.
    \item \textbf{Implementation Parameters ($\phi$)}: These variables govern hardware-specific configurations. For FPGAs, these include parallel factors and loop tiling factors; for GPUs, these include parameters like batch size. 
    \item \textbf{Quantization variables ($q$):} EDD treats the selection of quantization levels (e.g., 8-bit vs. 16-bit) as a sampling process within the differentiable flow, allowing the system to learn the optimal precision for each layer.
\end{itemize}
\end{enumerate}
\begin{figure}[!hbtp]
    \centering
    \includegraphics[width=\linewidth]{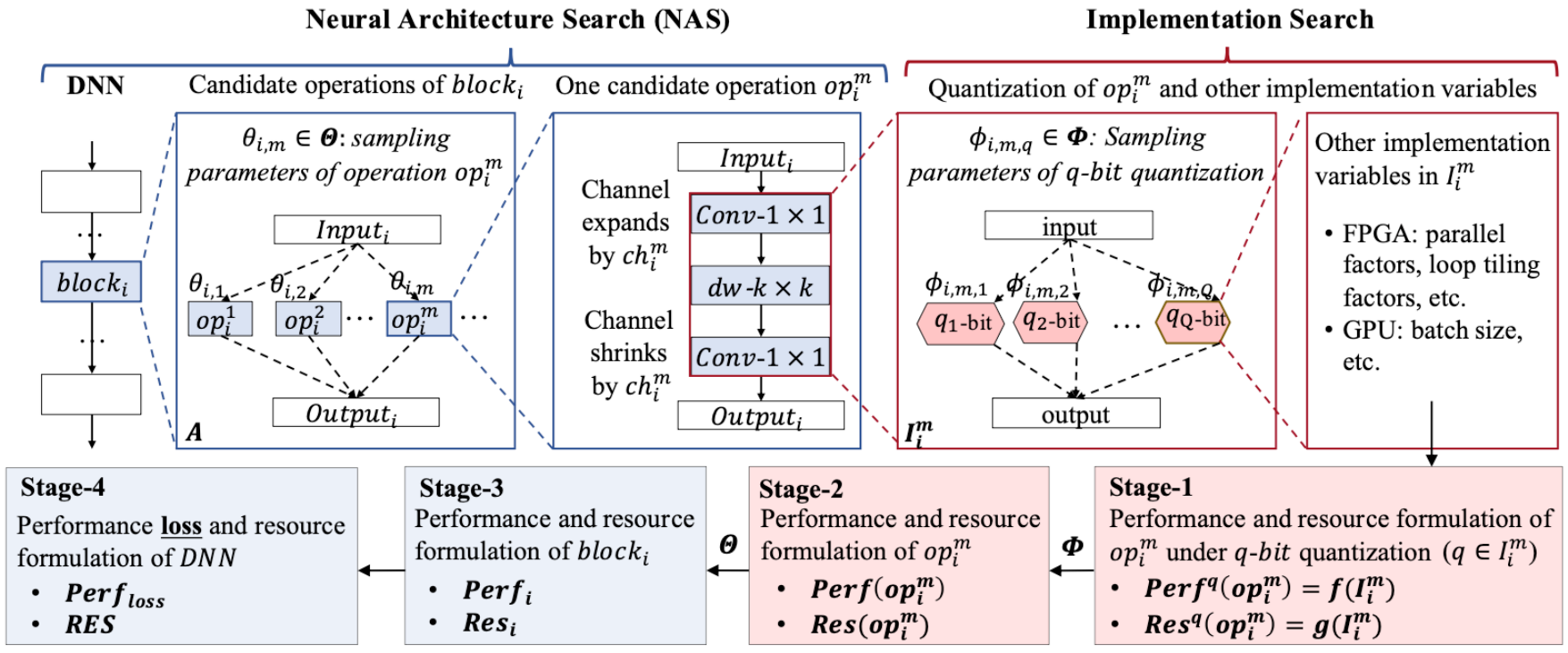}
    \caption{EDD \cite{li2020edd}, a four-staged differentiable NAS and Implementation Search flow. }
    \label{fig:edd}
\end{figure}
Figure \ref{fig:edd} illustrates EDD’s joint differentiable search space for neural architecture search and hardware implementation search. On the left, the NAS component selects, for each DNN block, one candidate operation from a set of MBConv-style choices using learnable architecture parameters $\theta$; each candidate operation specifies architectural features such as kernel size and channel expansion/shrink ratio. On the right, the implementation-search component assigns hardware-dependent variables to each operation, especially quantization precision through learnable parameters $\phi$, together with platform-specific variables such as FPGA parallelism and loop tiling or GPU batch size. The bottom row shows how EDD propagates these choices through a four-stage formulation: first estimating performance and resource cost for each operation under each quantization ($q$) setting, then aggregating over quantization choices, then aggregating over candidate operations within each block, and finally deriving the full DNN-level performance loss and resource usage. These differentiable estimates are inserted into the EDD objective, allowing architecture variables and implementation variables to be optimized simultaneously rather than treating hardware cost as a fixed post-search constraint.
\subsection{System Evaluation and Results}
The effectiveness of EDD is demonstrated across heterogeneous platforms, showing its ability to produce "friendly pairs" that outperform manually tuned or decoupled designs. 
\begin{itemize}
    \item \textbf{GPU Results (EDD-Net-1):} Targeting GPU platforms, EDD-Net-1 is searched to minimize latency while maintaining accuracy. It consistently outperforms state-of-the-art NAS works, achieving latency reductions of 1.4$\times$ to 2.0$\times$. 
    \item \textbf{Recursive FPGA Results (EDD-Net-2):} Evaluated using the CHaiDNN \cite{Xilinx2018CHaiDNN} framework, this configuration achieves the shortest latency recorded on FPGA, with speedups ranging from 1.1$\times$ to 1.53$\times$.
    \item \textbf{Pipelined FPGA Results (EDD-Net-3):} When compared against DNNBuilder~\cite{zhang2018dnnbuilder}, EDD-Net-3 achieves 1.45$\times$ higher throughput while simultaneously delivering higher accuracy.
\end{itemize}
\subsection{Conclusion of the EDD Framework}
The EDD methodology proves that systematic, differentiable co-design is superior to iterative, top-down approaches. By solving $\{A, I\}$ simultaneously using gradient descent, it eliminates the tedious engineering effort of balancing software and hardware metrics manually, providing a scalable path for future adaptive and evolvable AI systems.

\section{Hardware-Aware Co-design for Large Language Models}

\subsection{Challenges in LLM Inference}

The rapid scaling of large language models (LLMs) has fundamentally shifted the bottlenecks of AI system design. As model sizes grow exponentially, often increasing by an order of magnitude each year, both the number of parameters and the size of intermediate states have outpaced the improvements offered by traditional hardware scaling. Contemporary large language models, such as OpenAI's GPT-5 \cite{singh2026openaigpt5card}, comprise hundreds of billions to several trillions of parameters, placing unprecedented pressure on memory systems. 

Beyond model size, the auto-regressive decoding paradigm proposed in Transformer architecture further exacerbates inference inefficiency. During inference, each token generation requires a full pass of the model weights from High-Bandwidth Memory (HBM) to the compute units (SRAM/ALUs). The \textit{attention} \cite{vaswani2017attention} mechanism used by Transformer models cause per-token latency and memory footprint to quadratically increase with respect to the sequence length. As a remedy to this problem, Key-Value (KV) cache was proposed to store past token KV states, which grow linearly with respect to the sequence length. Although caching helps avoid redundant computations, as sequence lengths and batch sizes grow, exhausted HBM severely degrades the throughput of the system. Consequently, LLM inference becomes predominantly memory-bandwidth-bound rather than compute-bound, leading to severe under-utilization of modern accelerators’ arithmetic units. 

These characteristics make LLM inference difficult to parallelize and ill-suited to conventional acceleration techniques that focus primarily on increasing raw compute throughput. Applications requiring low-latency responses, such as interactive assistants and real-time decision systems, are particularly impacted. Addressing these challenges therefore requires hardware-aware algorithmic innovations that rethink the decoding process itself, rather than incremental optimizations to existing execution pipelines. 

\subsection{Medusa: Fast Decoding via A3C3-Inspired Co-design}
\begin{figure}[!htbp]
    \centering
    \begin{subfigure}[b]{0.45\linewidth}
        \centering
        \includegraphics[width=\linewidth]{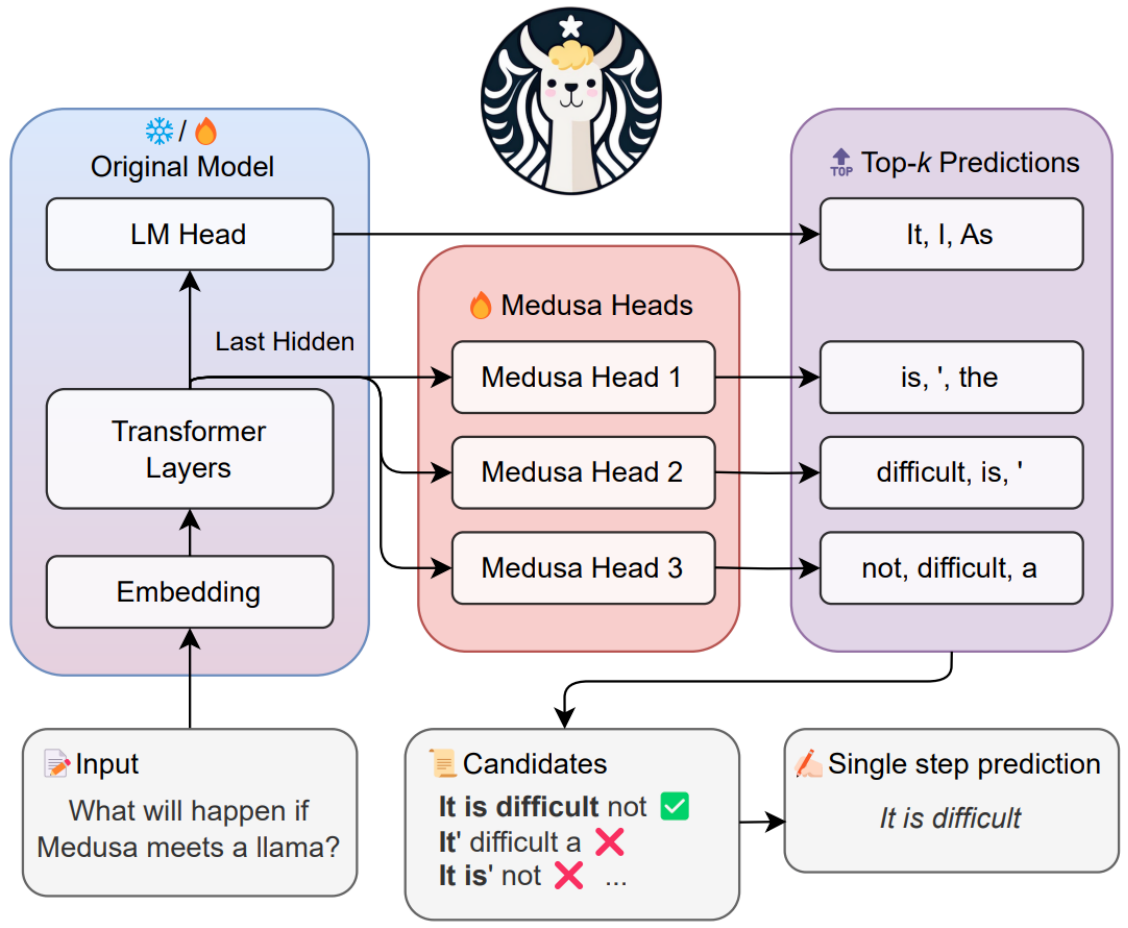}
        \caption{Medusa's parallel speculative decoding heads. Each head predicts a token at its corresponding future position given the base model's hidden states.}
        \label{fig:medusa-arch}
    \end{subfigure}
    \hfill
    \begin{subfigure}[b]{0.45\linewidth}
        \centering
        \includegraphics[width=\linewidth]{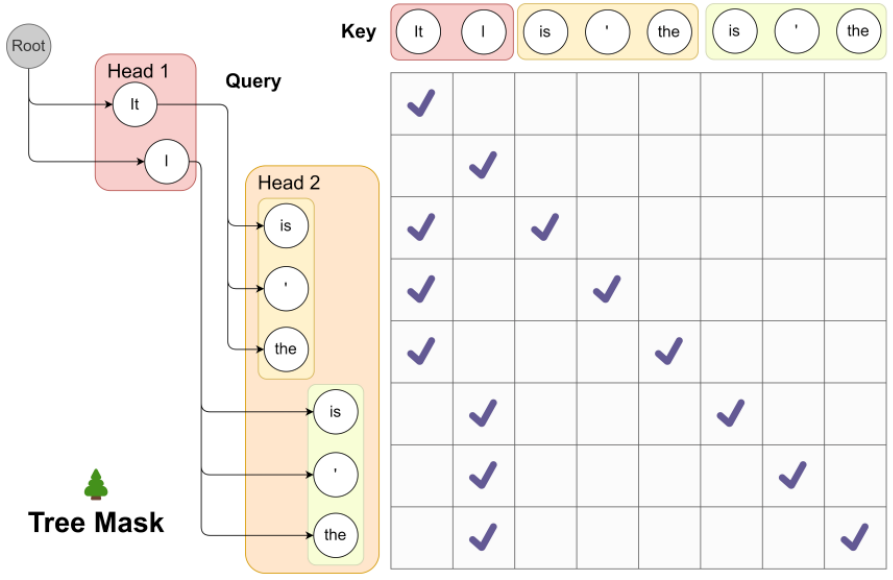}
        \caption{Tree-based attention mask proposed in Medusa for ensuring each token only attends its predecessors.}
        \label{fig:medusa-tree}
    \end{subfigure}
    \caption{Illustration of Medusa \cite{cai2024medusa} in A3C3 design.}
    \label{fig:medusa}
\end{figure}

Medusa addresses these inefficiencies by applying the A3C3 philosophy: co-designing the algorithmic decoding structure to match the parallel processing capabilities of modern accelerators. Instead of generating a single next token ($t+1$) per forward pass, Medusa predicts multiple future tokens ($t+1$, $t+2$, ..., $t+k-1$, $t+k$) simultaneously, where $k$ is the number of decoding heads, as shown in Figure \ref{fig:medusa-arch}.

\subsubsection{Parallel Speculative Decoding Heads and Medusa Tree Attention}
Medusa augments the original LLM backbone with multiple Medusa heads. Each head is a simple Feed-Forward Network (FFN) with a single residual connection, predicting the distribution of the $k$-th future token. To handle the inherent uncertainty of predicting multiple steps ahead, Medusa adopts a tree-based speculative decoding strategy:
\begin{itemize}
\item \textbf{Candidate Token Generation:} Each Medusa head proposes the top-$K$ most likely tokens for its respective position. These candidates are combined to form a set of speculative paths represented on a tree structure.
\item \textbf{Efficient Tree Attention:} To verify these paths in a single forward pass, Medusa uses a specialized attention mask. This mask ensures that a candidate token at position $t+2$ only attends to its valid predecessors in the tree (the tokens at $t$ and $t+1$) within the set of continuations (paths) it belongs to. This is illustrated in Figure \ref{fig:medusa-tree}.
\item \textbf{Token Acceptance Criteria:} To determine the validity of candidate sequences, a \textit{typical acceptance} criterion based on the model's predictive distribution is employed. This approach uses the original model's $(p)$ probability as a natural gauge for token quality, establishing a dynamic threshold for acceptance.

Specifically, given a context $\mathbf{x} = (x_1, x_2, \dots, x_n)$ and a candidate continuation
$(x_{n+1}, x_{n+2}, \dots, x_{n+k})$, the candidate token $x_{n+j}$ is accepted if its target-model probability exceeds an entropy-adaptive threshold:
\begin{align}
    p(x_{n+j} \mid x_1, \dots, x_{n+j-1})
    >
    \min \left(
        \epsilon,\,
        \delta \exp\left(
            -H\left(p(\cdot \mid x_1, \dots, x_{n+j-1})\right)
        \right)
    \right),
\end{align}
where:
\begin{itemize}
    \item $1 \leq j \leq k$.
    \item $p(\cdot \mid x_1, \dots, x_{n+j-1})$ denotes the target model's next-token distribution over the vocabulary $\mathcal{V}$.
    \item $p(x_{n+j} \mid x_1, \dots, x_{n+j-1})$ denotes the target-model probability assigned to the candidate token $x_{n+j}$.
    \item $H(\cdot)$ denotes the Shannon entropy of the target next-token distribution, defined as
    \begin{align}
        H(P) = -\sum_{v_i \in \mathcal{V}} p_i \log p_i,
        \qquad
        p_i = P(v_i) = p(v_i \mid x_1, \dots, x_{n+j-1}),
    \end{align}
    where $v_i$ is the $i$-th vocabulary token and $p_i$ is its probability under the target distribution $P$.
    \item $\epsilon$ denotes the hard cap on the acceptance threshold, preventing the threshold from exceeding a fixed maximum value.
    \item $\delta$ is a scaling factor for the entropy-dependent threshold.
    \item $\delta \exp(-H(P))$ defines the entropy-adaptive component of the threshold: low-entropy distributions yield a stricter threshold, while high-entropy distributions yield a lower threshold and allow more diverse candidate tokens to be accepted.
\end{itemize}
The final sequence for the current decoding step is determined by selecting the longest accepted prefix across all evaluated continuations.
\end{itemize}
\subsubsection{Training Methodology: Medusa-1 and Medusa-2}The co-search for the optimal Medusa configuration involves two distinct training regimes:
\begin{enumerate}
\item \textbf{Medusa-1 (Parameter-Efficient):} Only the Medusa heads are trained while the backbone remains frozen. This uses a self-distillation \cite{hinton2015distilling} loss, where the Medusa heads are trained to predict the original model's output distribution. This allows the system to gain speedups with minimal training overhead.
\item \textbf{Medusa-2 (Joint Co-optimization):} Both the backbone and the heads are fine-tuned together. This enhances the "predictability" of the backbone for the heads, yielding higher acceptance rates of speculative tokens and, consequently, higher effective parallelism.
\end{enumerate}
The training objective uses a cross-entropy loss summed across all $k$ heads: $\mathcal{L} = \sum_{k=1}^{K} \lambda_k \mathcal{L}_k$, where $\lambda_k$ is a weighting factor for the $k$-th future position.

\subsubsection{Experimental Results and Performance Analysis}
Medusa is evaluated across a range of model sizes and deployment settings to assess its effectiveness and scalability. Experiments were conducted using two training strategies: Medusa-1, which fine-tunes only the speculative decoding heads, and Medusa-2, which jointly fine-tunes both the Medusa heads and the LLM backbone. Across benchmarks, Medusa-2 consistently achieved higher speedups, demonstrating the benefits of end-to-end co-optimization. 

On Vicuna-7B and Vicuna-13B \cite{vicuna2023} models, Medusa achieved substantial reductions in decoding latency compared to baseline auto-regressive inference. Medusa delivered up to 3.6$\times$ speedup while maintaining output quality comparable to the baseline. With Vicuna-7B, Medusa-1 performed 2.2x acceleration while Medusa-2 performed 2.8x. Further evaluations across multiple LLM architectures show that Medusa’s performance gains scale with model size, as larger models suffer more acutely from memory-bound decoding and therefore benefit more from speculative parallelism. 

Hardware-level analysis revealed that Medusa significantly improves operational intensity during inference. By verifying multiple candidate tokens per forward pass, Medusa increases the ratio of useful floating-point operations to memory accesses, leading to better utilization of accelerator compute resources. In standard decoding, the GPU is underutilized because it waits for data transfer from memory. In Medusa, the target model verifies a tree-structured set of candidate tokens using parallel tree attention. Although the tree may contain many candidate nodes, e.g., 64 total candidates, its sequential verification cost is primarily determined by the tree depth, i.e., the number of token levels, rather than the tree width, i.e., the number of alternatives per level. Thus, a wide candidate tree can be processed in roughly the time required to process only a few sequential token positions, corresponding to the number of tree levels. Since the memory access cost for the backbone weights is the same, the "useful work" per byte transferred increases significantly. On an NVIDIA A100 GPU, this shift moves LLM inference closer to a compute-bound regime, addressing a fundamental inefficiency of conventional decoding pipelines. 

By co-designing the speculative decoding algorithm with hardware execution characteristics, Medusa substantially increases effective parallelism while preserving the correctness guarantees of auto-regressive generation. Importantly, this approach mitigates memory bandwidth pressure by amortizing the cost of model weight and KV cache accesses across multiple verified tokens. As a result, Medusa improves both compute utilization and end-to-end inference latency without decreasing the generation quality. 

\subsubsection{Broader Impact and Framework Adoption}

Medusa has been integrated into NVIDIA’s production inference stack through TensorRT-LLM \cite{nvidia_tensorrt_llm_speculative_decoding}, demonstrating its practical viability on modern accelerator platforms. NVIDIA highlights Medusa as a low-latency speculative decoding method for Llama 3.1 serving on HGX H200 \cite{nvidia_hgx_platform} systems, where eight H200 GPUs are connected through NVLink Switch to provide high-bandwidth multi-GPU execution, as reported in \cite{nvidia_medusa_hgx_h200_2024}. Instead of relying on a separate draft model, Medusa augments the original model with additional decoding heads that predict multiple future-token candidates, which are then validated in parallel by the base model using tree-based attention. This design reduces autoregressive bottlenecks while avoiding the deployment complexity and distribution mismatch of a separate drafter. In the reported TensorRT-LLM setup by NVIDIA, Medusa improves Llama 3.1 70B throughput from 184 to 268 tokens/s/user and Llama 3.1 405B throughput from 56 to 108 tokens/s/user, corresponding to over 1.5$\times$ and over 1.9$\times$ speedups, respectively, with bitwise-identical accuracy to the base model under their evaluation setting. These results show that Medusa is not only an algorithmic acceleration technique, but also a hardware-aware inference optimization that can exploit high-bandwidth GPU systems and mature serving frameworks to deliver meaningful latency reductions in large-scale LLM deployment.

\subsection{SnapKV: KV Cache Compression via A3C3 Co-design} 

As large language models are increasingly deployed on long-context tasks, inference efficiency is further constrained by the rapid growth of the key–value (KV) cache. During auto-regressive decoding, KV pairs for all past tokens must be stored and repeatedly accessed by the attention mechanism. For long prompts and extended generations, this cache can consume tens of gigabytes of memory, significantly increasing hardware requirements and limiting scalability. 

More critically, per-token decoding latency typically increases with prompt length, because each new token attends over all previously cached keys and values. As the KV cache grows, inference becomes increasingly memory-bandwidth bound, especially for long-document question answering (QA), multi-turn conversations~\cite{zheng2023judging}, and retrieval-augmented workloads with large effective context. While prior KV cache compression methods reduce memory usage, they often rely on static heuristics or fixed retention policies that do not adapt to input-specific attention dynamics, leading to accuracy degradation or limited effectiveness. 

These challenges highlight the need for a dynamic, context-aware KV compression mechanism that can significantly reduce memory footprint and decoding latency without compromising model quality. 

\subsubsection{SnapKV: Observation-Driven KV Cache Compression}

SnapKV~\cite{li2024snapkv} introduces a hardware-aware and fine-tuning-free KV cache compression framework based on a key empirical observation: during generation, the attention allocation patterns learned early in the prefill stage remain consistent across subsequent tokens. Specifically, SnapKV identifies an observation window, typically the most recent segment of the prompt, from which attention patterns can reliably predict the subset of prompt tokens that are most critical for generation. 

The SnapKV workflow has two stages. First, it designates a short observation window at the end of the prompt and computes attention from the tokens in this window to all earlier prompt tokens. Formally, let the prompt be $x_{1:N}$, with an observation window of length $W$,
\[
\mathcal{O}=\{N-W+1,\dots,N\},\qquad \mathcal{P}=\{1,\dots,N-W\}.
\]
For layer $\ell$ and head $h$, let $\mathbf{q}^{\ell,h}_t$ denote the query vector of observation token $t$, and let $\mathbf{k}^{\ell,h}_i$ denote the key vector of prompt token $i$, where $d$ is the per-head hidden dimension. SnapKV computes the attention from each observation token $t \in \mathcal{O}$ to earlier prompt tokens $i \in \mathcal{P}$ as
\[
a^{\ell,h}_{t,i}
=
\mathrm{softmax}_{i\in\mathcal{P}}
\left(
\frac{
\left\langle \mathbf{q}^{\ell,h}_t, \mathbf{k}^{\ell,h}_i \right\rangle
}{\sqrt{d}}
\right).
\]
It then aggregates these attention weights across the observation window tokens to obtain an importance score for the prompt token:
\[
s^{\ell,h}_i=\sum_{t\in\mathcal{O}} a^{\ell,h}_{t,i},\qquad i\in\mathcal{P}.
\]
The top-ranked prompt tokens are selected as critical, and only their corresponding key and value states are retained in the compressed KV cache. Although the final selection is still made over individual prompt positions $i \in \mathcal{P}$, SnapKV can optionally apply local pooling to the attention scores before the Top-$K$ selection. This pooling step smooths the importance scores across nearby positions, allowing neighboring tokens to influence a token’s final importance score. Therefore, neighboring tokens may affect which positions enter the Top-$K$ set, but SnapKV does not automatically retain all neighbors; it retains only the key/value states of the final selected Top-$K$ tokens.

\[
\tilde{s}^{\ell,h}_i=\max_{j\in[i-r,\ i+r]\cap \mathcal{P}} s^{\ell,h}_j \quad (\text{or average pooling}),
\]
and selects critical tokens under a budget $K$:
\[
\mathcal{I}^{\ell,h}=\mathrm{TopK}\!\left(\tilde{s}^{\ell,h},\,K\right)\subseteq\mathcal{P}.
\]
This process yields an attention allocation pattern that captures the most informative tokens for the downstream generation process, which can be expressed as a binary mask
\[
m^{\ell,h}_i=\mathbf{1}\{i\in\mathcal{I}^{\ell,h}\},\qquad i\in\mathcal{P}.
\]

Second, SnapKV applies concatenation to compress the KV cache. Selected critical tokens are pooled and concatenated with the observation window tokens to form a compact representation of the original prompt. Specifically, for each layer $\ell$ and head $h$, SnapKV forms the compressed KV cache by
\[
\mathbf{K}^{\ell,h}_{\mathrm{keep}}
=\big[\mathbf{K}^{\ell,h}_{\mathcal{I}^{\ell,h}};\ \mathbf{K}^{\ell,h}_{\mathcal{O}}\big],
\qquad
\mathbf{V}^{\ell,h}_{\mathrm{keep}}
=\big[\mathbf{V}^{\ell,h}_{\mathcal{I}^{\ell,h}};\ \mathbf{V}^{\ell,h}_{\mathcal{O}}\big].
\]
During decoding, attention is computed only over this compressed KV cache, substantially reducing memory access and computation costs. Importantly, because SnapKV derives compression decisions from model-internal attention patterns computed for each input prompt, it dynamically adapts to different prompts and instructions, avoiding the limitations of static compression strategies.

\subsubsection{Experimental Results and Scalability}

SnapKV is evaluated on a diverse set of long-context benchmarks to assess both efficiency gains and quality preservation. In experiments using the LWM-Text-Chat-1M long-context chat model, SnapKV reports up to a 3.6$\times$ decoding speedup at a 16K-token sequence length with batch size two. At the same time, KV-cache memory consumption is reduced by up to 8.2$\times$, significantly lowering the GPU memory required for long-context inference.

Further experiments demonstrate SnapKV’s robustness across a range of prompt lengths and datasets. On LongBench~\cite{bai2024longbench}, SnapKV maintains accuracy close to the baseline while operating with a substantially compressed KV cache. In extreme long-context demonstrations, the authors show SnapKV enabling inference with up to 380K context tokens on a single A100-80GB GPU under their experimental setup, an input length that would typically be infeasible with an uncompressed KV cache under the same memory budget.

SnapKV is also evaluated on a needle-in-a-haystack task, which measures a model’s ability to retrieve fine-grained information embedded deep within long documents. Even when the KV cache is compressed to as few as 1024 retained tokens—in configurations supporting contexts up to 380,000 tokens, SnapKV achieves negligible degradation in retrieval performance, indicating that the observation-driven compression mechanism preserves the critical context needed for accurate retrieval.

Collectively, these results support SnapKV as a practical, hardware-oriented approach for long-context LLM inference. By deriving input-dependent KV retention decisions from attention behavior, SnapKV  reduces KV-cache memory footprint and per-token decoding cost while maintaining generation quality, enabling more scalable deployment on memory-constrained and latency-sensitive platforms.

\section{Discussion and Future Directions}

The A3C3 methodology demonstrates that co-design is not a niche optimization technique, but a general paradigm for building efficient AI systems across model families, application domains, and hardware platforms. The systems discussed in this chapter show that the dominant bottleneck changes significantly across deployment regimes: edge vision is constrained by power, latency, spatial resolution, and embedded resource budgets; differentiable architecture search is constrained by the combinatorial explosion of algorithmic and implementation choices; large language model inference is constrained by autoregressive decoding, memory bandwidth, and KV-cache growth. Despite these differences, the same principle recurs across all cases: substantial gains arise when the algorithm, memory behavior, and accelerator implementation are optimized as a coupled system rather than as independent layers.

SkyNet illustrates the importance of co-design in resource-constrained edge vision. In UAV-based object detection, the challenge is not merely to reduce the number of parameters, but to preserve detection accuracy for small objects while satisfying strict power and latency constraints on embedded FPGA and GPU platforms. SkyNet addresses this by jointly considering neural building blocks, pooling positions, channel expansion and shuffling, feature-map bypassing, quantization, and accelerator-level dataflow. This demonstrates that a compact model is not necessarily hardware-efficient by default; the network structure must expose the right computation patterns and memory-access behavior for the target device. The broader lesson from SkyNet is that edge AI requires architectures that are hardware-friendly by construction, especially when the application imposes simultaneous constraints on accuracy, throughput, power and form factor.

EDD extends this idea by replacing coarse, heuristic search with differentiable architecture and implementation co-search. While SkyNet uses a staged exploration process with Particle Swarm Optimization, EDD formulates the neural architecture and hardware implementation as a merged search space. This transition is important because the design space of modern AI systems is too large to explore effectively through manual tuning or greedy heuristic search alone. By relaxing architecture choices, quantization decisions, and hardware mapping parameters into differentiable variables, EDD enables gradient-based optimization over a unified objective that accounts for accuracy loss, performance loss, and resource constraints. This provides a more scalable foundation for A3C3 because it allows the search process to reason directly about the interaction between model topology and hardware execution cost.

The LLM case studies show that A3C3 remains relevant even when the bottleneck shifts from compute-bound to memory-bound autoregressive inference. In large language models, the dominant inefficiency is not simply insufficient compute throughput, but poor utilization of available compute due to sequential decoding and repeated data movement. Medusa addresses this issue by modifying the decoding algorithm itself: multiple speculative heads generate candidate future tokens, and tree-based attention verifies several candidate paths in a single forward pass. This increases the amount of useful computation performed per model-weight access and improves accelerator utilization. From the A3C3 perspective, Medusa is important because it shows that hardware-aware co-design can occur at the inference-algorithm level, not only at the model-architecture or accelerator-dataflow level.

SnapKV further broadens the scope of A3C3 by focusing on the memory footprint and access cost of long-context inference. As context lengths grow, the KV cache becomes a central system bottleneck, limiting batch size, increasing memory traffic, and raising per-token latency. SnapKV addresses this problem through an observation-driven compression mechanism that uses attention behavior from a short prompt window to identify the most important cached tokens. Unlike static compression policies, this method adapts to each input prompt and preserves the context most relevant to downstream generation. Its importance lies in showing that co-design must include memory budget and data retention policies, not only arithmetic acceleration. For future long-context models, deciding what information to store, move, compress, or discard will be as important as optimizing the neural computation itself.

Collectively, these systems suggest that future AI accelerators should not be designed around fixed assumptions about model structure. Instead, accelerators and models should be developed through adaptive co-search processes that expose tunable parameters at multiple levels: network topology, operator selection, precision, scheduling, memory layout, cache retention, decoding strategy, and interconnect communication. A mature A3C3 framework should therefore support hierarchical optimization, where local operator-level decisions are coordinated with global system-level objectives such as end-to-end latency, throughput, energy efficiency, memory capacity, and quality-of-result. This is particularly important as AI workloads become increasingly heterogeneous, combining dense computation, sparse attention, retrieval, multimodal processing, and long-context reasoning within a single serving pipeline.

One important future direction is distributed accelerator co-design. Current co-design methods often focus on a single device or a tightly coupled accelerator platform, but frontier AI systems increasingly operate across multiple GPUs, nodes, memory tiers, and storage systems. In this regime, the design problem expands beyond selecting efficient operators or model blocks. It must also consider tensor parallelism, pipeline parallelism, expert parallelism, KV-cache placement, interconnect bandwidth, communication overlap, and scheduling across heterogeneous memory and accelerator devices. Extending A3C3 to distributed inference would allow the search process to jointly optimize model partitioning, communication patterns, memory placement, and decoding algorithms. This is especially relevant for large language models and mixture-of-experts systems, where communication and memory movement can dominate execution time.

A second direction is heterogeneous accelerator co-design. Future AI platforms are unlikely to rely on a single type of processor. Instead, they may combine GPUs, FPGAs, ASICs, CPUs, near-memory accelerators, and domain-specific inference engines. Each device type offers different strengths: GPUs provide high-throughput dense computation, FPGAs offer reconfigurable dataflow and low-latency execution, ASICs provide energy efficiency for fixed workloads, and CPUs remain useful for orchestration and irregular control flow. A3C3 can provide a principled framework for mapping different parts of the AI workload to the most appropriate hardware substrate. For example, dense matrix operations may remain on GPUs, while KV-cache compression, token routing, retrieval, or sparsity-aware scheduling may be offloaded to specialized units. This type of heterogeneous co-design requires the search objective to incorporate not only per-device performance, but also data movement, synchronization, programmability, and runtime adaptability.

A third direction is memory-centric co-design. The SkyNet, Medusa, and SnapKV case studies all point to the same broader trend: memory movement is increasingly the limiting factor in practical AI deployment. In edge vision, memory bandwidth and on-chip buffer size constrain the feasible feature-map resolution and layer structure. In LLM decoding, repeated weight loading and KV-cache access limit arithmetic utilization. In long-context inference, the KV cache itself becomes a dominant memory object. Future A3C3 methodologies should therefore treat memory as a first-class design dimension. This includes co-optimizing model structure with cache hierarchy, compression policy, prefetching strategy, memory placement, and data reuse. Rather than designing accelerators only around peak compute throughput, future systems should optimize the full data lifecycle of AI inference.

A fourth direction is dynamic and input-adaptive co-design. Many current co-design methods produce a static model-accelerator pair that is optimized for an average workload or benchmark distribution. However, real-world AI workloads are highly variable. Images may differ in object density and resolution; prompts may differ in length, entropy, and retrieval requirements; LLM outputs may vary in predictability across decoding steps. Methods such as Medusa and SnapKV already suggest that runtime behavior can be exploited to improve efficiency: speculative decoding benefits when future tokens are predictable, and KV compression benefits when attention patterns reveal a small set of critical tokens. Future A3C3 systems should extend this idea by dynamically adapting precision, sparsity, cache size, decoding depth, parallelism, and accelerator scheduling based on input difficulty and inference-time signals, such as prediction entropy, token acceptance rate, model uncertainty or predictive confidence.

A fifth direction is quality-aware co-design for generative AI. Traditional hardware optimization often focuses on latency, throughput, energy, and resource utilization, while treating model quality as a fixed constraint. Generative AI complicates this view because quality is multidimensional and sometimes difficult to measure. For LLMs, preserving generation quality may involve factuality, instruction following, reasoning consistency, diversity, and user preference alignment. For image and video generation, quality may involve visual fidelity, semantic alignment, temporal consistency, and distributional realism. Future A3C3 frameworks must therefore incorporate richer QoR metrics into the search objective. This is particularly important for approximate inference techniques such as speculative decoding, KV compression, sparsity, quantization, and early exiting, where small algorithmic changes can have subtle effects on output quality.

A sixth direction is the integration of A3C3 with emerging computing technologies. Near-memory computing, processing-in-memory, chiplet-based accelerators, optical interconnects, analog computing, and non-volatile memory systems all create new design opportunities, but also introduce new constraints. These technologies often require models and algorithms to be restructured around their physical characteristics, such as limited precision, restricted data movement, analog noise, memory endurance, or communication topology. A3C3 is well suited to this setting because it does not assume that the algorithm is fixed before hardware design begins. Instead, it can search for algorithmic structures that naturally exploit the strengths and tolerate the limitations of emerging hardware.

Finally, future A3C3 research should address automation, transferability, and usability. A key promise of co-design is to reduce manual engineering effort, but practical deployment still requires accurate cost models, reliable search spaces, compiler support, and integration with production inference frameworks. Search results must transfer across datasets, hardware generations, batch sizes, and deployment conditions. They must also be interpretable enough for system designers to trust and maintain. Building reusable co-design libraries, standardized hardware-aware benchmarks, and end-to-end toolchains will therefore be essential for making A3C3 practical beyond isolated case studies.

In summary, the systems discussed in this chapter show a clear progression in the scope of co-design. SkyNet demonstrates co-design for compact edge vision; EDD formalizes differentiable architecture and implementation co-search; Medusa applies hardware-aware algorithmic restructuring to LLM decoding; and SnapKV extends co-design to memory retention and long-context inference. Together, they indicate that future efficient AI systems will not emerge from optimizing models or accelerators independently. They will require cross-layer methodologies that jointly reason about algorithmic structure, hardware execution, memory movement, runtime adaptivity, and application-level quality constraints. A3C3 provides a foundation for this direction and offers a path toward sustainable, scalable, and hardware-aware AI deployment.

\section{Conclusion}

This chapter has presented A3C3 paradigm as a comprehensive methodology for AI algorithm and accelerator co-design, co-search, and co-generation. Through systems such as SkyNet, EDD, Medusa, and SnapKV, A3C3 has been shown to deliver substantial improvements in deep learning performance, efficiency, and scalability. As AI models and hardware platforms continue to evolve, co-design will remain a central principle for achieving sustainable and efficient AI systems. 

\section{Acknowledgements}
The authors gratefully acknowledge the researchers whose foundational work forms the intellectual basis of this book chapter. We extend our sincere appreciation to the authors of SkyNet, EDD, Medusa, and SnapKV for their pioneering contributions to algorithm-accelerator co-design, differentiable architecture search, and efficient large language model inference. This work builds directly on the collective insights of these research teams. Additionally, this work was supported by the AMD Center of Excellence and the IBM-Illinois Discovery Accelerator Institute. The authors gratefully acknowledge their support and collaboration in advancing research on hardware-aware AI systems.
\bibliographystyle{plain}
\bibliography{references}
\end{document}